\documentclass[twoside,a4paper,12pt]{article}  

\usepackage{indentfirst}  
\usepackage{bm}    
\usepackage{graphicx}  
\usepackage{subfig}

\begin{document}    




\begin{center}
\LARGE\bf Chaos control in random Boolean networks by reducing mean damage percolation rate$^*$   
\end{center}


\begin{center}
\rm Nan Jiang$^{\rm 1\dag}$ \ and  \ Shijian Chen$^{\rm 2**}$
\end{center}

\begin{center}
\begin{footnotesize} 
\begin{sl}
$^{\rm 1}$ Department of Automation, Tsinghua University\\
Beijing, China, 100084\\
\end{sl}   
\begin{sl}
$^{\rm 2}$ Graduate University of Chinese Academy of Sciences\\
Beijing, China, 100049\\
\end{sl}  
\end{footnotesize}
\end{center}

\footnotetext{\hspace*{-.45cm}\footnotesize $^*$This work has been supported in part by the NNSF of China under Grants 60874018,60736022 and 60821091.}
\footnotetext{\hspace*{-.45cm}\footnotesize $^\dag$Corresponding author. E-mail: jn07@mails.tsinghua.edu.cn}
\footnotetext{\hspace*{-.45cm}\footnotesize $^{**}$E-mail: chenshijian@amss.ac.cn}

\vspace*{2mm}

\begin{center}
\begin{minipage}{15.5cm}
\parindent 20pt\footnotesize
Chaos control in Random Boolean networks is implemented by freezing part of the network to drive it from chaotic to ordered phase. However, controlled nodes are only viewed as passive blocks to prevent perturbation spread. This paper proposes a new control method in which controlled nodes can exert an active impact on the network. Controlled nodes and frozen values are deliberately selected according to the information of connection and Boolean functions. Simulation results show that the number of nodes needed to achieve control is largely reduced compared to previous method. Theoretical analysis is also given to estimate the least fraction of nodes needed to achieve control.
\end{minipage}
\end{center}

\begin{center}
\begin{minipage}{15.5cm}
\begin{minipage}[t]{2.3cm}{\bf Keywords:}\end{minipage}
\begin{minipage}[t]{13.1cm}
Random Boolean networks; Chaos control; Mean damage percolation rate
\end{minipage}\par\vglue8pt
\end{minipage}
\end{center}

\section{Introduction}  
Random Boolean networks (RBNs) are an abstract model of genetic regulatory networks suggested by Kauffman \cite{Kau}, promising to reveal some principles of living systems with a systematic view. RBNs' application has been found in areas like sociology, neural networks and music generation \cite{Intro,Yeast,Memory}.

The classical model of RBN is described as follows. There are $N$ nodes, every node with $K$ edges pointed from others. The state of every node takes Boolean value. State of a node in the next time step is determined by the states of its input nodes and an individual Boolean function (a lookup rule table). The probability for values in the lookup rule tables to take $1$ is called the \emph{bias} of the RBN, marked as $p$. Many other models are proposed considering different updating schemes, while in this paper we mainly focus on the classical model.

As specific networks can be extremely large and complex, studies concentrate on the behavior of certain ensembles of RBNs. For example, $N$, $K$ and $p$ are often fixed, and connections, rule tables and initial states supposed to be randomly picked at the beginning of revolution. With some stochastic approach, theoretical analysis can be given to such ensembles and verified by simulation with randomly generated RBNs. Then, the properties found in an ensemble can be applied to a particular system which is a member of it.

Similar to other dynamic systems, phases of \emph{ordered}, \emph{chaotic} and \emph{critical} can be found in RBNs, usually defined by properties of perturbation spread. The critical condition is one of the most important issues in RBNs. Another interesting topic is chaos control. A network in chaos phase may be driven into periodic behavior when states of a certain percentage of nodes are determined externally \cite{Control,Signal,Hong}. In previous work, nodes to be controlled and their frozen values are all picked up randomly. Our work shows that they can be deliberately selected to achieve high control efficiency.

The rest of the paper is organized as follows: Section 2 introduces related work on RBN control. Section 3 explains the concept of canalizing and the basic idea of our method. Section 4 elaborates the method and gives theoretical analysis. Simulation results are given in Section 5. Section 6 makes conclusion and discuss about future work.

\section{Related work on RBN control}  
We first define some symbols about RBN. Denote the state of node $i$ at time t as $x_i(t)$, and the Boolean function of node i as $f_i$. The input nodes of node $i$, namely the nodes pointing to node $i$, are denoted as $i_1$,\dots,$i_K$ . RBN evolves in the way
\begin{equation}
x_i(t+1) = f_i(x_{i_1}(t),\dots,x_{i_K}(t)), i=1,2,\dots,N.
\end{equation}

Ref.\cite{Signal} gives a general form of RBN control. Define $\gamma^{max}$ as the maximum percentage of controlled nodes. $F_\tau(t)$ is any positive function with periodic $\tau$, varying between $0$ and $1$. Let $\gamma(t)=\gamma^{max}F_\tau(t)$, denoting the percentage of controlled nodes at time $t$. Before evolution, $N\gamma^{max}$ controlled nodes are picked up randomly from the network. For convenience we assume that their sequence numbers are $1,\dots,N\gamma^{max}$. Each controlled node is given a random value $C_i\in \{0,1\}, i=1,\dots,N\gamma^{max}$. At each time step $t$, the states of node $1,\dots,N\gamma^{max}$ are fixed to their corresponding $C_i$, while the remaining controlled nodes are updated freely. For RBNs in chaos phase, a $\gamma^{max}$ large enough can drive the networks into ordered phase, which usually appear as periodic behavior with an integer multiple of $\tau$ to be the period.

To estimate the least $\gamma^{max}$ needed to control the networks, an approximation method is used \cite{Phase}. The basic idea is to assume that the state of one of the nodes flips, which introduces perturbation (or damage) into the network. If the state of a successor node flips at the next time step, we say the perturbation spreads (or percolates) to this successor node. If the mean probability that one damage spreads in a control period $\tau$ is larger than 1, damage amplifies in the network, meaning the system is in chaos phase, and vice versa. Thus, letting the mean probability to be 1, we can get the critical condition of the system, corresponding to the least $\gamma^{max}$ to achieve control. The critical condition is
\begin{equation}
[2p(1-p)K]^\tau\prod_{t=0}^{\tau-1}(1-\gamma(t))=1.
\end{equation}
This is because the probability for a free node to flip is $p$, and a node has $K(1-\gamma(t))$ free successor nodes in average. Letting $\gamma^{max}=0$, we get the critical condition of free RBN:
\begin{equation}
2p(1-p)K=1.
\end{equation}

In this analysis, controlled nodes only serve as passive blocks to prevent damage spread. However, as the driven power of the network, we expect they have active impact on the network. Moreover, controlled nodes and their values are chosen totally randomly. It can be expected that better efficiency be achieved if more information about the network is utilized in selection. In the following sections, the paper will show that this can be done by selecting the controlled nodes and their fixed values according to a certain measure. Before that, we need to introduce the concept of canalizing.

\section{Canalizing and the basic idea of our method}
A Boolean function is canalizing if, whenever one input node (canalizing input node) takes a certain value (canalizing value), the function always gives out the same output \cite{Canal}. For example, $x_1 or f(x_2,\dots,x_K)$ is a canalizing function because the function always yields 1 when $x_1=1$.

Our basic idea is: if a controlled node is a canalizing input node to one of its free successors, and it is fixed to the canalizing value, the successor node always yield the same output and is therefore insensitive to any perturbation. In this way, fixing a controlled node does not only prevent itself from spreading damage, but does the same thing to at least one free node, changing it into a controlled node equivalently.

As each node has several successor nodes, a controlled node's value should be fixed to the very value which is canalizing to more successor nodes. All nodes in the network can also be sorted according to such a measure. Those who freeze most free nodes should be selected as controlled nodes so that high control efficiency is achieved.

We extend the concept of canalizing to a more general idea. Let the percentage of 1s in node $i$'s rule table be $R_i$. When one of $i$'s input nodes, node $j$, is fixed to a constant value $C$, denote the percentage of 1s in $i$'s possible output values in the rule table as $R_i|_{x_j=C}$. Using these symbols, canalizing can be expressed as $R_i|_{x_j=C}=0$ or 1. If $R_i$ is close to 0 or 1, the node can also prevent damage spread efficiently as different inputs are likely to yield a same output (we will prove this in the next section).

To fix a controlled node $j$ to a certain value $C$ means its successor node $i$'s percentage of 1s in possible output values changes from $R_i$ to $R_i|_{x_j=C}$. Counting the canalizing successor nodes, which is mentioned above, does not take $R_i|_{x_j=C}$ that is close to 0 or 1 into account, and ignore the possible negative effect on successor nodes as $R_i|_{x_j=C}$ may be more close to 0.5 than $R_i$ is. Based on this observation, we suggest our method which uses another measure instead of counting canalizing successor nodes.

\section{RBN control based on reducing mean damage percolation rate}
In eq.(3), mean damage percolation rate is calculated by $2p(1-p)$, which depends on a priori parameter $p$, the ``genotype'' of the network. For a given RBN, $p$ can be unknown and mean damage percolation rate should be estimated by $R_i$, the ``phenotype'' of the network.

We adopt the same assumption that eq.(3) uses: at first the input of a node is randomly picked, and a flip of state in a certain input node makes the input changes randomly to another one. For node $i$, there are $2^KR_i$ 1s and $2^K(1-R_i)$ 0s in the rule table, thus the probability for node $i$ to flip is
\begin{equation}
R_i\frac{2^K(1-R_i)}{2^K-1}+(1-R_i)\frac{2^KR_i}{2^K-1}=\frac{2^{K+1}}{2^K-1}R_i(1-R_i).
\end{equation}
As $R_i(1-R_i)$ is a quadratic function with a maximum at $R_i=0.5$, it proves the statement in Section 3 that nodes with $R_i$ close to 0 or 1 can prevent damage spread effectively. Since damage percolation rate of a single node is obtained, the mean damage percolation rate of the network should be
\begin{equation}
\frac{1}{N}\frac{2^{K+1}}{2^K-1}\sum_{l=1}^{N}R_l(1-R_l).
\end{equation}
On the other hand, we calculate the expectation of $\displaystyle\frac{1}{N}\sum_{l=1}^{N}R_l(1-R_l)$:
\begin{equation}
E(\frac{1}{N}\sum_{l=1}^{N}R_l(1-R_l))=E(R_i(1-R_i)=p-p^2+Var(R_i).
\end{equation}
As $R_i$ is Bernoulli distributed,
\begin{equation}
Var(R_i)=\frac{p(1-p)}{2^K}.
\end{equation}
Thus we have,
\begin{equation}
E(\frac{1}{N}\sum_{l=1}^{N}R_l(1-R_l))=\frac{2^K-1}{2^K}p(1-p)
\end{equation}
which shows that $\displaystyle\frac{1}{N}\frac{2^{K+1}}{2^K-1}\sum_{l=1}^{N}R_l(1-R_l)$ can be an estimation of $2p(1-p)$.

From two different points of view, we reach the same conclusion that $\displaystyle\frac{1}{N}\frac{2^{K+1}}{2^K-1}\sum_{l=1}^{N}R_l(1-R_l)$ is the mean damage percolation rate of the network, which is also proved as a theorem in Ref.\cite{Sensi}. If it is reduced as much as possible, the network is likely to fall into ordered phase.

We implement chaos control in RBNs with deliberate selection of controlled nodes and their fixed values as follows: let $O(i)$ be the set of output nodes of node $i$. Define
\begin{equation}
S_{i,C}=R_i(1-R_i)+\sum_{j\in O(i)}[R_j(1-R_j)-R_j|_{x_i=C}(1-R_j|_{x_i=C})]
\end{equation}
and
\begin{equation}
S_i=max\{S_{i,0},S_{i,1}\},
\end{equation}
\begin{eqnarray}
C_i = \left\{
\begin{array}{l l}
0 & \quad \mbox{if }S_{i,0}>S_{i,1}\\
1 & \quad \mbox{otherwise}\\
\end{array} \right..
\end{eqnarray}

Sort all nodes according to their $S_i$ and a new order $(1),\dots,(N)$ is obtained so that $S_{(i)}\geq S_{(j)}$ for any $i<j$. At every time step, use node $(1),\dots,(N\gamma(t))$ as controlled nodes and fix them to $C_{(1)},\dots,C_{(N)}$.

When $x_i$ is fixed to $C$, $R_i(1-R_i)$ is removed from $\displaystyle\sum_{l=1}^{N}R_l(1-R_l)$ and damage percolation rate of all node $i$'s output nodes are altered. $S_{i,C}$ is in proportion to the amount that the mean damage percolation rate of all rest nodes is reduced by fixing $x_i=C$. As $C$ can be 0 or 1, we choose the fixed value which makes more reduction. We select the nodes with largest $S_i$ to be controlled nodes so that the mean damage percolation rate of the network is reduced to least. The mean damage percolation rate at time $t$ is
\begin{equation}
\frac{\displaystyle\frac{2^{K+1}}{2^K-1}(\displaystyle\sum_{l=1}^{N}R_l(1-R_l)-\sum_{i=1}^{N\gamma(t)})S_{(i)})}{N(1-\gamma(t))}.
\end{equation}
And the critical condition is
\begin{equation}
\prod_{t=0}^{\tau-1}(\frac{\displaystyle\frac{2^{K+1}}{2^K-1}(\displaystyle\sum_{l=1}^{N}R_l(1-R_l)-\sum_{i=1}^{N\gamma(t)}S_{(i)})}{N})=1.
\end{equation}

We should mention that such analysis does not take into account the situation that a free node has two or more controlled nodes as input nodes. However, this is rare when $K\gamma(t)<1$, which is ignored here.

To estimate the least $\gamma^{max}$ needed to implement control is difficult because the out-degrees of the nodes vary, as the out-degree of a single node has a standard deviation of nearly $\sqrt{k}$. Therefore, the distribution of $S_i$ is hard to obtain. In addition, $S_i$ has strong correlation with each other, which makes it difficult to calculate the expectation of $S_{(i)}$.

We give a rough estimation of $\gamma^{max}$ by adopting the following assumptions: (1) $S_{i,C}$ of different $i$ are independent with each other, and so are the out-degrees. (2) Consider $S_{i,C}$ and out-degrees of a certain network which exactly fits the presumed distribution, and calculate $\displaystyle\sum_{i=1}^{N\gamma(t)}S_{(i)}$ of such a network as an estimation. We explain our assumptions below in detail.

Let $O_i=\#O(i)$ and $O$ be a random variable independent identically distributed (i.i.d.) with $O_i$. $O$ is Bernoulli distributed with parameter $(N-1,\displaystyle\frac{K}{N-1})$. As $\displaystyle(N-1)\frac{K}{N-1}=K$, when $N$ is very large, $O$ is approximately Poisson distributed with $\lambda=K$, which means
\begin{equation}
P(O=n)\approx \frac{K^n}{n!}e^{-K}.
\end{equation}
Assumption (2) means, we consider a network which has $P(O=n)\cdot N$ nodes with out-degree $n$.

Let $X_i=R_i(1-R_i)$. For $j\in O(i)$, $X_{i,j,C}=R_j(1-R_j)-R_j|_{x_i=C}(1-R_j|_{x_i=C})$. The meaning of assumption (1) is, $X_{i,j,C}$ of different $i$,$j$ are independent with each other and $X_i$. However, $X_{i,j,0}$ and $X_{i,j,1}$ are not independent because both of them rely merely on a same rule table. Here we have
\begin{equation}
S_{i,C}=X_i+\sum_{j\in O(i)}X_{i,j,C}.
\end{equation}
For $O_i=n$, let $S_{O=n,C}$ be an random variable i.i.d. with $S_{i,C}$. Let $X_{C=0}$ and $X_{C=1}$ be two random variables dependent on a same rule table, i.i.d. with $X_{i,j,0}$ and $X_{i,j,1}$. Since the distribution of $X_{C=0}$ and $X_{C=1}$ can be obtained by enumerating possible rule tables of a node, the distribution of $S_{O=n,C}$ can also be obtained by iteration
\begin{eqnarray}
& P(S_{O=n+1,0}=s_0,S_{O=n+1,1}=s_1) \nonumber \\
&\displaystyle =\sum_{\theta_0,\theta_1}P(S_{O=n,0}=s_0-\theta_0,S_{O=n,1}=s_1-\theta_1)P(X_{C=0}=\theta_0,X_{C=1}=\theta_1).
\end{eqnarray}
For $O_i=n$, let $S_{O=n}$ a random variable i.i.d. with $S_i$. Its distribution can be obtained by
\begin{eqnarray}
& P(S_{O=n}=s) \nonumber \\
& =\displaystyle\sum_{s>s'}P(S_{O=n,0}=s,S_{O=n,1}=s')+\displaystyle\sum_{s\geq s'}P(S_{O=n,0}=s',S_{O=n,1}=s).
\end{eqnarray}
In the very network we consider, the number of nodes whose $S_i=s$ is approximately
\begin{equation}
\sum_{n=0}^{N-1}P(O=n)P(S_{O=n}=s)N.
\end{equation}

So far, we are able to estimate
$\displaystyle\sum_{i=1}^{N\gamma(t)}S_{(i)}$ with assumption (2).
For $K=3$, $p=0.5$ and $K=3$, $p=0.3$, we calculate the estimation
of $\displaystyle\sum_{i=1}^{m}S_{(i)}, m=1,2,\dots,N$ and that of a
real network, shown in Fig.1. For normalization, the X and Y axes
are both divided by $N$. It can be seen that the estimation is
accurate, especially when $m$ is not large.

\vspace*{4mm}
\begin{figure}
\centering \subfloat[Part
1][p=0.5]{\includegraphics[scale=0.7]{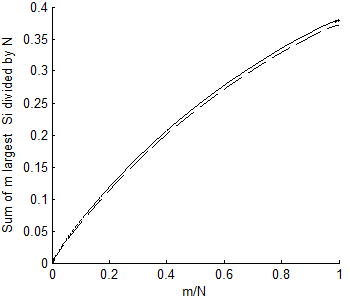}
} \subfloat[Part 2][p=0.7]{\includegraphics[scale=0.7]{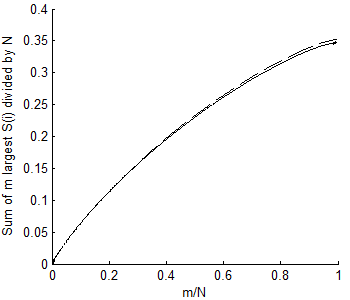} }\\
\caption{Curves of sum of $m$ largest $S_i$ are given. (a) is drawn
under $p=0.5$ and (b) is under $p=0.3$. The solid line is
theoretical estimation, and the dashed line is calculated from a
real network.}
\end{figure}

\section{Simulation results}
To compare our method with previous work, we perform 100 simulations for each $(p,\gamma^{max})$ with $N=1000$, $K=3$. $F_\tau(t)=sin^2(\pi t/\tau)$ and $\tau=50$. We count within how many of 100 simulations the network is driven into periodic behavior and plot the fractions on Fig.2. If 100 simulations all succeed in control, the small rectangle at $(p,\gamma^{max})$ is painted white. If they all fail, the rectangle is painted black. Intermediate fractions are painted according to the color bar shown at the right of the plots. As can be seen, the least $\gamma^{max}$ needed to implement control is largely reduced. Theoretical bounds of the two methods are given by the dot lines.
\vspace*{4mm}


\begin{figure}
\centering \subfloat[Method in
Ref.{\cite{Signal}}]{\includegraphics[scale=0.7]{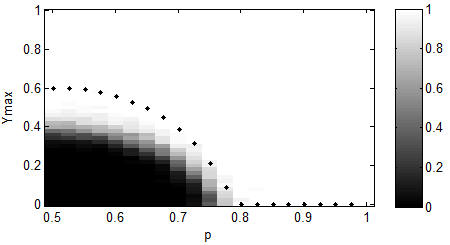}
} \subfloat[Our method]{\includegraphics[scale=0.7]{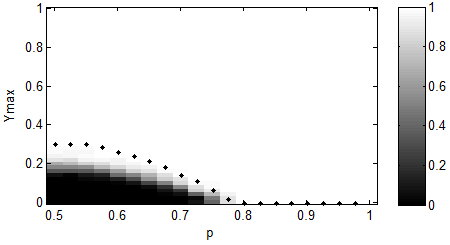} }\\
\caption{Simulation result and theoretical bound of two methods.}
\end{figure}

\section{Conclusion and future work}
In this paper, we propose a new control method in RBNs. Inspired by the concept of canalizing, we use the controlled nodes not only as passive blocks to prevent damage spread, but to exert active influence on the network by altering the property of its free nodes. Then we develop a measure with more generality and clearer physical meaning than counting canalizing nodes, and propose our method based on such a measure. Besides controlled nodes, the method determines the fixed values. The innovation of this method is that we make full use of connection information. More precise measure could be used to replace mean damage percolation rate suggested in this paper, while the idea that fixing node has an influence on successor nodes could be remained to form a new method. However, this idea also results in the difficulties in estimating the least $\gamma^{max}$ needed to achieve control. Independence assumptions are adopted to give a rough estimation.

For future work, more information could be utilized to realize efficient control. Although we have already made most of connection and Boolean functions information, our control method fails to consider the dynamic states of the system, which means it is an open-loop control method. A measure of percolation rate such as sensitivity \cite{Sensi} which is related to RBN's current state could be used in selecting fixed values which vary with time, implementing a close-loop control. In addition, selection of controlled nodes should consider the topology of the network. Fixing specific nodes to build blocks could separate sensitive nodes apart to prevent damage spread cascading \cite{Doyle}. Finally, math tool is to be found to accurately analyze the method that connection information usage causes difficulty in calculating expectation of strong correlated sorted statistics.



\begin{thebibliography}{8}
\itemsep=-4pt plus.2pt minus.2pt  
\small
\bibitem{Kau} Kauffman S A 1969 \textit{Journal of Theoretical Biology} {\bf 22} 437

\bibitem{Intro} Gershenson C 2004 \textit{Workshop and Tutorial Proceedings, Ninth International Conference on the Simulation and Synthesis of Living Systems (ALife IX)} (MA) p160

\bibitem{Yeast} Li F T and Jia X 2006 \textit{Chinese Phys. Lett.} {\bf 23} 2307

\bibitem{Memory} Huang Z G, Wu Z X, Guan J Y and Wang Y H 2006 \textit{Chinese Phys. Lett.} {\bf 23} 3119

\bibitem{Control} Luque B and Sol\'{e} R V 1997 \textit{Europhys. Lett.} {\bf 37(9)} 597

\bibitem{Signal} Ballesteros F J and Luque B 2002 \textit{Physica A} {\bf 313} 289

\bibitem{Hong} Hong Y G and Xu X R 2010 \textit{Proceedings of the 29th Chinese Control Conference} (Beijing) p805

\bibitem{Phase} Luque B and Sol\'{e} R V 1997 \textit{Physical Review E} {\bf 55(1)} 257

\bibitem{Canal} Moreira A A and Amaral L A 2005 \textit{Physical Review Letters} {\bf 94(21)} 218702

\bibitem{Sensi} Schober S and Bossert M 2007 arXiv:0704.0197v1

\bibitem{Doyle} Zhou T, Carlson J M and Doyle J 2002 \textit{Proc. Natl. Acad. Sci.} {\bf 99(4)} 2049


\end{thebibliography}
\end{document}